\documentclass[useAMS,usenatbib]{mn2e}
\usepackage{graphicx}
\usepackage{color}
\newcommand{\kms}{$\mathrm{km\, s^{-1}\, }$}
\newcommand{\msun}{M_{\odot}}

\title[Stellar winds near massive black holes]{Stellar winds near massive black holes: \\ The case of the S-stars.}
\author[N. L\"utzgendorf et al.]{N. L\"utzgendorf $^{1}$\thanks{E-mail:
email@address (AVR); otheremail@otheraddress (ANO)}, E. van der Helm$^{2}$, I. Pelupessy$^{2,3}$ and S. Portegies Zwart$^{2}$ \\
$^{1}$ESA, Space Science Department, Keplerlaan 1, NL-2200 AG Noordwijk, The Netherlands\\
$^{2}$Leiden Observatory, Leiden University, PO Box 9513, NL-2300 RA, Leiden, The Netherlands\\
$^{3}$Institute for Marine and Atmospheric Research Utrecht, Department of Physics and Astronomy Utrecht,\\
 University Princetonplein 5, 3584 CC Utrecht, The Netherlands}
\begin{document}

\date{Accepted 2015 December 09. Received 2015 December 09; in original form 2015 September 18}

\pagerange{\pageref{firstpage}--\pageref{lastpage}} \pubyear{2002}

\maketitle

\label{firstpage}

\begin{abstract}
The Galactic center provides a unique laboratory to study the interaction of a supermassive black hole (SMBH) with its gaseous and stellar environment. Simulations to determine the accretion of stellar winds from the surrounding O-stars onto the black hole have been performed earlier, but in those the presence of the S-star system was ignored. The S-stars are a group of young massive B-stars in relatively close orbits around the black hole. Here we simulate those stars in order to study their contribution to the accretion rate, without taking the more distant and massive O-stars into account.  We use the Astrophysical Multi-purpose Software Environment (AMUSE) to combine gravitational physics, stellar evolution and hydrodynamics in a single simulation of the S-stars orbiting the supermassive black hole, and use this framework to determine the amount of gas that is accreted onto the black hole. We find that the accretion rate is sensitive to the wind properties of the S-stars (rate of mass-loss and terminal velocity). Our simulations are consistent with the observed accretion rate of the black hole only if the stars exhibit high wind massloss rates that are comparable with those of evolved 7-10 Myr old stars with masses of $M=19-25 \, M_{\odot}$. This is in contrast with observations that have shown that these stars are rather young, main-sequence B-stars. We therefore conclude that the S-stars cannot account for the accretion rate alone.% However, the shock heated gas from colliding stellar winds of the population of S-stars may be responsible for the recently observed outflows of Sgr  A$\ast$.}

\end{abstract}

\section{Introduction}\label{sec:intro}

Within the central tenth of a parsec in the middle of our galaxy there is a concentration of young stars that interact with a supermassive black hole - the S-stars \citep[e.g.][]{schoedel_2002,ghez_2003,eisenhauer_2005, ghez_2005}. In contrast to gas, the orbits of the stars are governed by gravitation only and therefore provide an excellent tracer for the gravitational potential in our Galactic center. This unique setup provides the best measurement of the mass of a black hole to date and unambiguously confirms the existence of a supermassive black hole in the center of our galaxy \citep[][]{genzel_2010}.

\citet{schoedel_2002} and \citet{ghez_2003} carried out high resolution near-IR imaging and spectroscopy of the central region of our Milky Way. They found a star (nowadays known as S2) closely bound to a central massive object with  an orbital period of 15.2 years \cite[see also][]{gillessen_2009b}. The monitoring of its orbit for over 10 years provided the best observation of a Keplerian orbit around a massive black hole to date. Today, the number of well-determined orbits has risen from 1 to 28 while 109 stars in total are still being monitored \citep{gillessen_2009}.

The S-stars in the galactic center are thought to be massive, early B-type stars and therefore should exhibit hot stellar winds \citep[e.g.][]{kudritzki_2000}. These winds provide gaseous material that can be accreted by the black hole which is thought to be the source of X-ray emission close to Sgr A$^\ast$. \citet{baganoff_2003} used Chandra X-ray spectroscopic imaging of the central parsec of the galaxy to measure the Bondi accretion rate of the central black hole. They found that the X-ray emission at the position of Sgr A$^\ast$ is extended and consistent with the accretion radius for a $\sim 10^6 \, M_{\odot}$ black hole ($\sim 12 000$ au). Furthermore, they estimated the Bondi accretion rate to be $\sim 10^{-6} \, M_{\odot}$ year$^{-1}$ at the Bondi radius. This is in contrast with constraints from linear polarization measurements at 150 GHz and above which limit the accretion rate near the event horizon to $\sim 10^{-7} - 10^{-8}  \, M_{\odot}$~year$^{-1}$ \cite[e.g.][]{bower_2003}. However, the latter estimates strongly depend on the assumption of equipartition between particle and magnetic field energy of a uniform magnetic field.

\citet{nayakshin_2005} estimated the accretion rate of Sgr. A$^\ast$ by looking at the pericenter passage of S2 in the year 2002. The absence of a noticeable change in the steady emission of Sgr A$^\ast$ suggest less dense accretion flows and constrains the accretion rates to $\sim 10^{-7} \, M_{\odot}$ year$^{-1}$. We note, however, that this model is based on a specific accretion flow model by \citet{yuan_2002} and might not be generally valid.

\citet{perets_2009} simulated the evolution of the orbital eccentricities of the S-stars using N-body simulations and found that only 20-30\% of the S-stars are tidally disrupted by the central supermassive black hole and that they are not likely to be ejected as hypervelocity stars  by close encounters with stellar mass black holes. This suggests that the winds of the S-stars could have provided accretion material for the supermassive black hole for the last 10 - 100  Myr.

A full simulation of wind accretion onto Sgr  A$^\ast$ was performed by \citet{cuadra_2006}. In this work they studied the accretion of slow and fast winds originating from young massive O-stars in the vicinity of the supermassive black hole out to a radius of 1 pc. In their most sophisticated models, they put stellar wind sources on realistic orbits around Sgr A$^\ast$, and included slow winds ($v_w \sim 300$ \kms) as well as radiative cooling. They found that compared to models with stationary wind emitting stars, the accretion rate decreases by an order of magnitude for wind sources following circular orbits, and fluctuates by $\sim 50 \%$. In \citet{cuadra_2008} these models were updated with observational data on orbits and wind properties on the individual stars. With the updated orbits and non-zero eccentricities they found the accretion to be dominated by a few close-by stars with slow winds ($v_w \le 750 $ \kms) and no obvious disk-like structure.

To fit observational X-ray surface brightness profile from Chandra observations \citet{shcherbakov_2010} developed a two-temperature radial inflow-outflow model for the radiation near Sgr A$^\ast$. They report an accretion rate of $\dot{M} = 6 \times 10^{-8} M_{\odot}$ year$^{-1}$ at a stagnation radius (position of zero of velocity stagnation point) $r_{st} = 1''.01 (\sim 8000$ au) and a temperature at the stagnation radius of $T_{st} = 3.2 \times 10^7$ K. These simulations and estimates, however, cover a region much larger than the S-star system and can therefore only be used as an approximate guideline to compare the results of this study.

The young massive O-stars in the vicinity of the black hole are not the only source of gas produced by stellar wind. The S-Stars are main sequence B-type stars which produce less wind but are located closer to Sgr. A$^\ast$. \citet{loeb_2004} were the first to consider the wind of the S-stars as a possible source for the accretion luminosity or at least accretion variability of Sgr. A$^\ast$. Using the assumption that gas particles move on Keplerian orbits, they concluded that the S-stars could provide the gas needed to explain the accretion rate for mass loss rates around $\sim 10^{-6} \, M_{\odot}$~year$^{-1}$.

In this work we simulate the motion of the S-stars and the wind emission/accretion onto the central black hole. By combining N-body dynamics, stellar evolution and hydrodynamics we investigate the influence of hot stellar wind on the gaseous vicinity of a massive black hole that is surrounded by a number of stars on bound orbits around the black hole. Furthermore, we investigate the temperature distribution of the gas produced by the wind and the Bondi accretion rate onto the black hole. In Section \ref{sec:sim} we describe the simulations and the individual codes it is composed of. Section \ref{sec:res} outlines the results of the various simulations and tests that were performed to verify the credibility of the code and in \ref{sec:obs} we compare our results to observational data. A summary and our conclusions are formulated in Section \ref{sec:concl}.

\section{Simulations}\label{sec:sim}

To simulate the accretion of stellar winds onto a central black hole it is required to combine three types of physics into one simulation, stellar evolution, hydrodynamics and gravitational dynamics. In order to do this, we use the Astrophysical Multi-purpose Software Environment \cite[][amusecode.org]{portegies_zwart_2009,portegies_zwart_2013,pelupessy_2013}. AMUSE is a software framework for astrophysical simulations, in which existing codes from different domains, such as stellar dynamics, stellar evolution, hydrodynamics and radiative transfer can be easily coupled. AMUSE uses Python to interface with existing numerical codes. The AMUSE interface handles unit conversions, provides a consistent object oriented interface, manages the state of the underlying simulation codes and provides transparent distributed computing. 
%It is a community effort with the main development by the AMUSE team at Leiden Observatory. AMUSE currently contains 56 codes among which there are 17 gravitational codes, 5 hydrodynamic codes, 5 stellar evolution codes, and 3 radiative transfer codes. For a further documentation and an overview of AMUSE we refer to \citet{portegies_zwart_2009, pelupessy_2013} and the AMUSE homepage amusecode.org. 
We have used AMUSE to combine a gravitational N-body system containing a massive black hole in the center with a Smooth Particle Hydrodynamics (SPH) code in order to simulate the behavior of stellar wind from massive stars near a massive black hole. The unique environment of AMUSE allows us to combine these codes with stellar evolution in order to obtain the ideal wind parameters and mass-loss rates for each star. Furthermore it allows a detailed analysis of temperature and density evolution in this environment and the accretion of stellar wind onto the central black hole. Our code solves the hydrodynamics and self-gravity of the gas, stellar evolution and N-body dynamics of the galactic center self-consistently. However, for the simulations presented here the most important processes are the response of the gas to the stellar and black-hole potential and the hydrodynamic response of the gas, since we only consider a limited time span.

\begin{figure}
  \centering  % this centres figure in column
  \includegraphics[width=0.99\columnwidth]{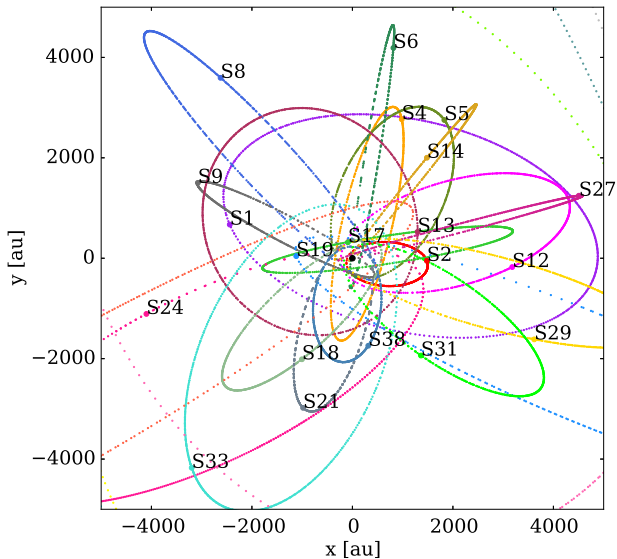}
  \caption{Projected integrated orbits of the innermost S-stars as they orbit the central black hole. The initial conditions of the code are taken from \citet{gillessen_2009}. For the integration of the orbits the MUSE internal gravitational code ph4 was used.}
  \label{fig:orbits}
\end{figure}

\subsection{Initial Setup}

The orbital parameters of the S-stars are well constrained due to the multiple observations and monitoring of the central regions in our galaxy. In order to start with the most realistic initial conditions we take the orbital parameters of the S-stars as reported by \citet{gillessen_2009}. Using multiple epochs of proper motions in combination with line-of-sight velocities the authors measured the orbits of the 27 innermost stars and derived their orbital properties from dynamical modeling. Using the Kepler equations, the orbital parameters are corrected to Cartesian positions and velocities adopting a central black hole of $M_{\bullet} = 4.45 \times 10^6 M_{\odot}$. In Figure \ref{fig:orbits} we show the orbits of the innermost S-stars from their initial conditions projected on the plane of the sky. All of the stars were assigned the same mass of $\sim 20 M_{\odot}$ and age of $\sim 9$ Myr. The mass loss rate calculated with the stellar evolution model for these parameters is $10^{-5}  M_{\odot}/$yr and it is the same for each star.

\subsection{Stellar Evolution}

The AMUSE framework offers multiple stellar evolution codes ranging from semi-analytic implementations such as \verb|SSE| \citep{hurley_2000} and  \verb|SeBa| \citep{ portegies_zwart_1996} to 1D Henyey stellar evolution codes such as \verb| EVTwin | \citep{eggleton_1971,glebbeek_2008} and \verb|MESA| \citep[][]{paxton_2011}. For our purposes, no complicated processes need to be considered for the stellar evolution as only mass-loss and effective temperature are important. Therefore, we use the more simple and less computational expensive code \verb|SSE|. By changing one line in the script, one could use more precise stellar evolution models, but the computational cost would go up drastically.

\verb|SSE| uses the polynomial fits as described in \citet{hurley_2000} to approximate the evolution of stars for a wide range of stellar parameters from zero-age main sequence to remnant stages. It also provides information about mass-loss at each time step that is used to generate the gas particles for the stellar wind. The code was used to simulate the evolutionary path of B-type stars with a metallicity of $Z  = 0.02$ as the galactic center is thought to be roughly solar \citep[][and references therein]{maness_2007} .

\subsection{Gravitational integration}

We simultaneously integrate the individual orbits of each star while evolving hydrodynamics and stellar evolution. We use \verb|ph4| \citep{mcmillan_2012} for the gravitational solver which provides a parallel and GPU-accelerated N-body integration module with block time steps \citep[using the Sapporo - GPU library][]{gaburov_2012} using a 4th-order Hermite integration scheme \citep{makino_1992}. In order to couple the gravitational N-body code to the gas particles we use the AMUSE \verb|bridge|. The \verb|bridge| module provides a self-consistent coupling between different components of a self-gravitating system using operator splitting \citep[see][]{fujii_2007,pelupessy_2012, pelupessy_2013}. 
In short, with bridge the gas and stars are integrated in each other's mutual gravitational field. The integration is realized via the 2nd order Verlet method which is applied every synchronization time step. In this way a system composed of multiple components is evolved self-consistently, taking the self gravity of the system as a whole into account, while choosing the most appropriate integrator for the self-gravity for each of the component subsystems.

\subsection{Hydrodynamics}

A crucial part of the simulation is the hydrodynamics of the gas that is produced from the stellar wind. For this, we use the particle based SPH code \verb|Fi| \citep{pelupessy_2009} that solves the Euler equations for the gas dynamics using a Galilean invariant Lagrangian method. The hydrodynamics are based on a particle representation of the fluid \citep{monaghan_1992}, in the conservative formulation of \citet{springel_2002}. The thermal evolution of the gas is either determined by an adiabatic equation of state, or by solving the heating/cooling balance of the gas, using far-UV heating and atomic line cooling \citep{wolfire_1995,pelupessy_2005}. For the latter we have two options: the internal cooling/heating of the code itself,  or an external cooling routine that is directly executed in the AMUSE script and uses an approximation of the cooling curves. This allows us to exchange \verb|Fi| with \verb|GADGET| \citep{springel_2005} another SPH particle based code in AMUSE. Switches in the code allow the user to decide between internal, external and no cooling of the gas. The gas is modeled with a metallicity of $Z=0.02$ and a dynamical smoothing length. We chose the time steps of the hydrodynamics, gravitational dynamics and stellar evolution to be one year. The kicks from \verb|bridge| are calculated every half time step (i.e., every $0.5 $ yrs).

\subsection{Stellar wind}

The stellar wind is simulated using the new AMUSE module \verb|stellar_wind| (van der Helm et al., in prep.) in which one can choose several wind models: The simple model contains a wind mechanism where, according to the mass loss given by the stellar evolution, SPH particles are released near the star with a temperature of $0.8 T_{\mathrm{eff}}$ and a radial terminal wind velocity calculated from the stellar parameters using Equations (8), (9) and (10) in \citet[][]{kudritzki_2000}. The other models simulate stellar wind in special circumstances like close to the stars or in dense clustered environments. 

For our purposes the first method with the simple wind model is enough to describe the system of S-stars since the closest distance to the central black hole is $\sim 100$ au for S14 which is large compared to the radius of the star of $\sim 4$ au. The wind is released in each time step after evaluating the mass-loss through stellar evolution. External/internal cooling is applied during each half step of the hydrodynamic simulation.

\subsection{Accretion onto the black hole}

\begin{figure*}
  \centering  % this centres figure in column
     \includegraphics[width=0.32\textwidth]{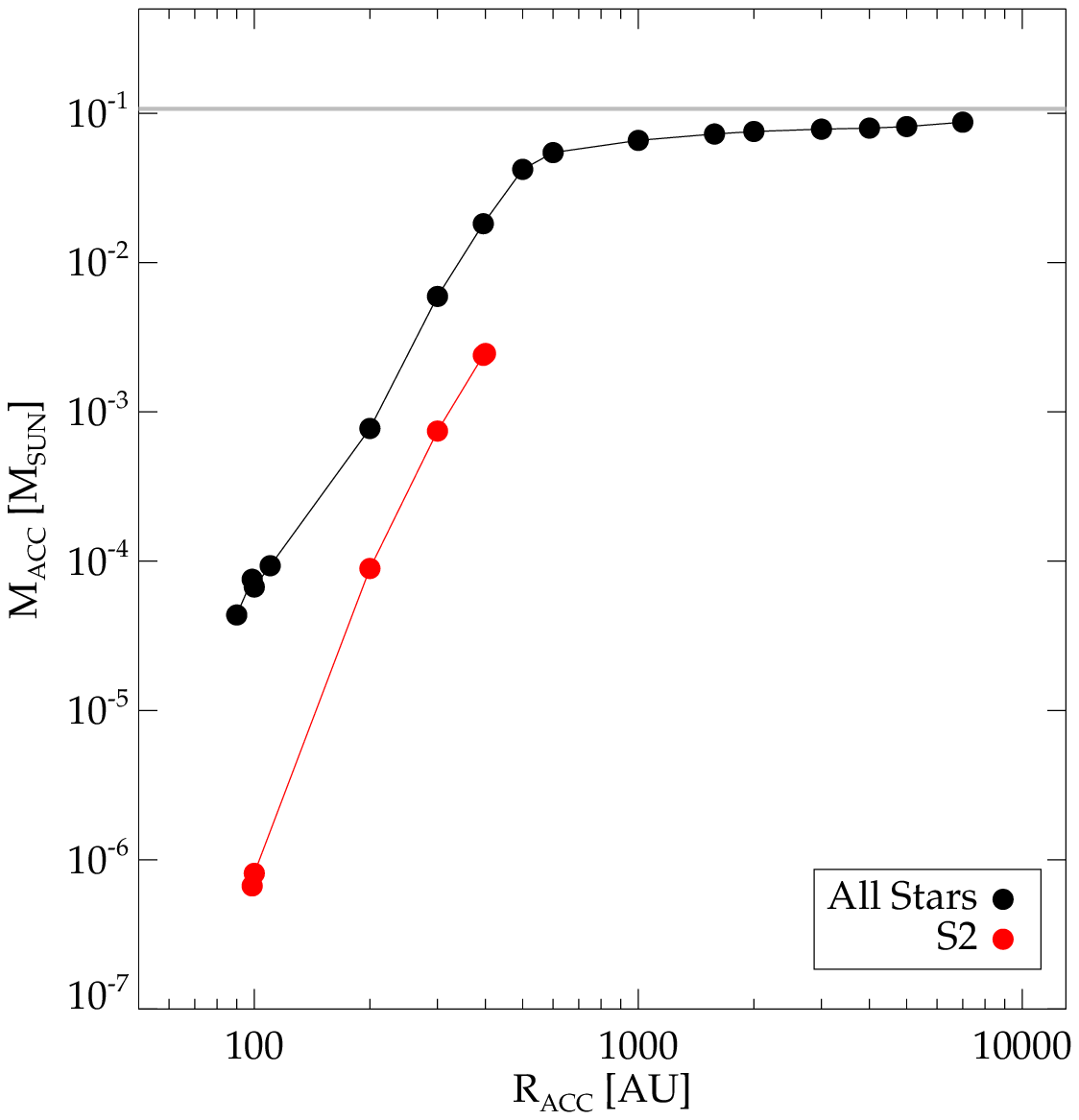}
     \includegraphics[width=0.32\textwidth]{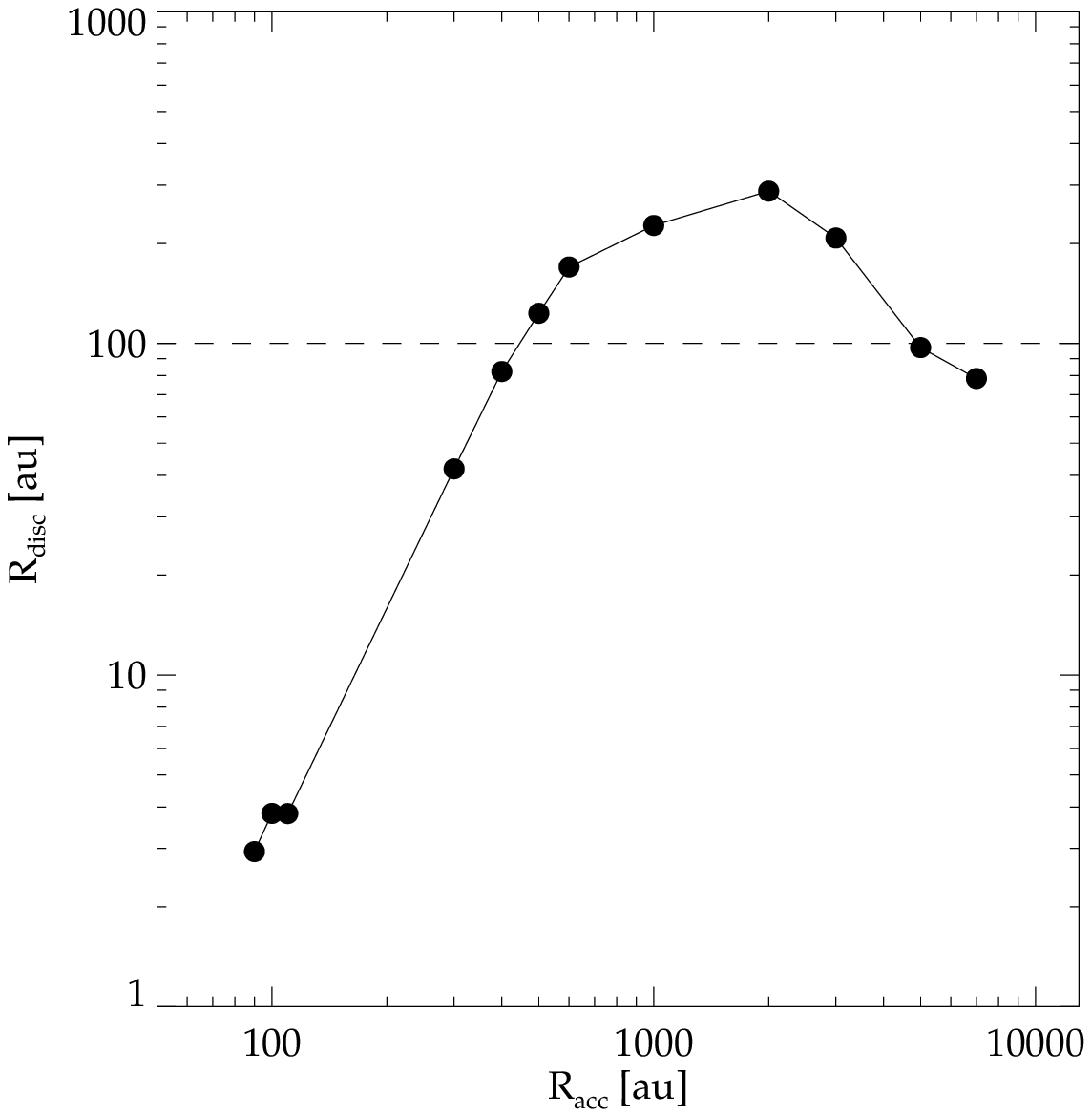}
     \includegraphics[width=0.32\textwidth]{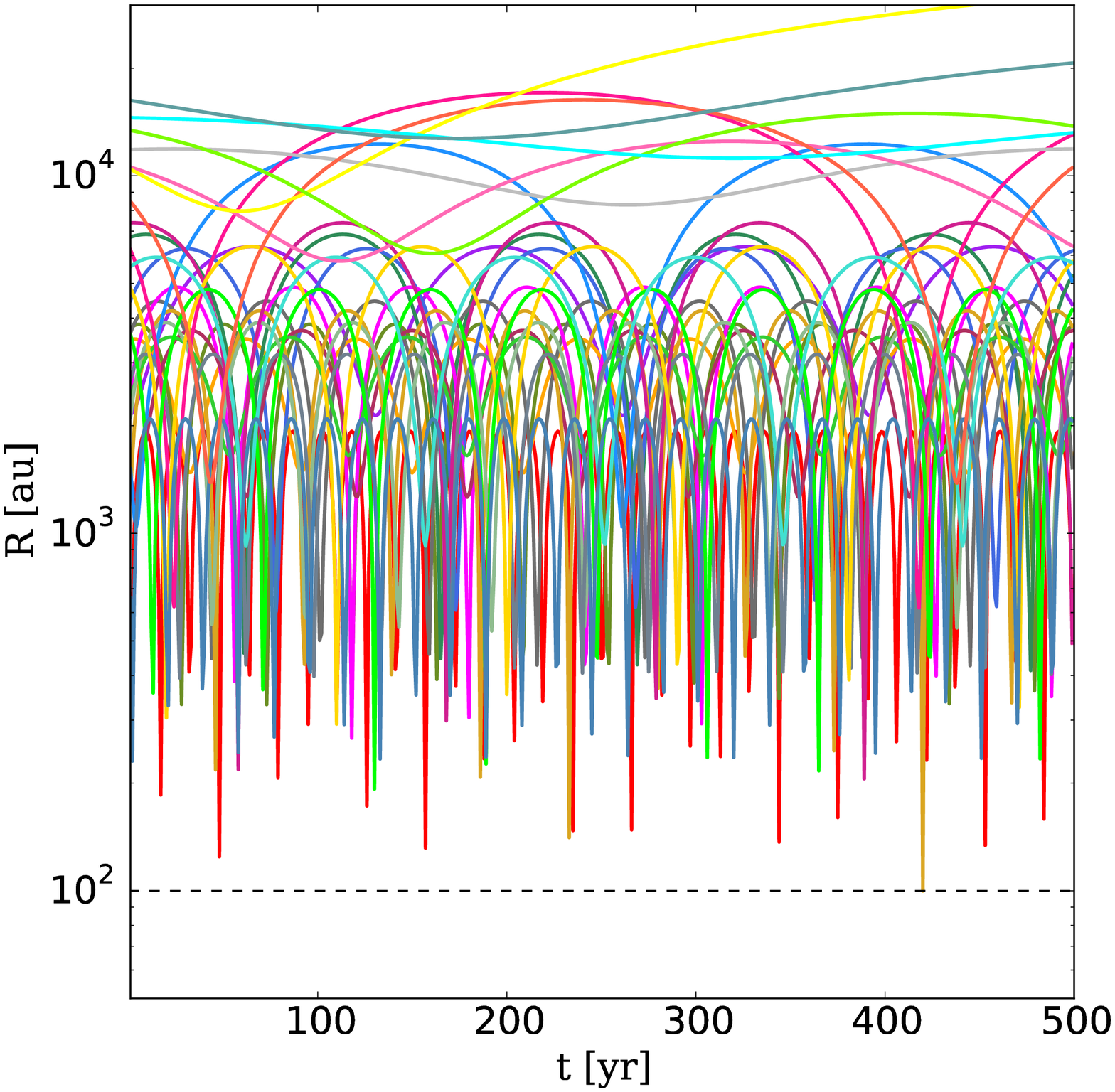}
     
  \caption{Determining the accretion radius. Left: total accreted mass after 500 years as a function of accretion radius. The grey line marks the sum of the gas available in the simulation. The curve for S2 does not extend further than its semi-major axis (and all of the gas is accreted). Middle: Size of possible accretion disc calculated from the mass and angular momentum inside the sink particle as a function of accretion radius. Right: Distance to the central black hole as a function of time for each star. The dashed line marks the largest sink radius that excludes all stars in their closest approach. }
  \label{fig:racc}
\end{figure*}

To simulate accretion we make use of the concept of a sink particle. Any gas particles that pass within the accretion radius of the sink particle are considered to be accreted. Their mass, linear and angular momenta are added to those of the sink particle. The accretion radius, or size of the sink particle, is a sensitive parameter in our simulations. By varying this radius up to 7000 au (which includes almost all orbits) the accretion rate varies strongly. In the left panel of Figure \ref{fig:racc} we show the total accreted mass after 500 years versus the accretion radius for the simulation of all the stars and the specific case of S2. The figure shows a clear trend of accretion rate with accretion radius and raises concern about the steepness of the slope. It only flattens when almost all available gas (upper line) is accreted. 

In order to constrain the accretion radius of the sink particle in our simulation we compute the size of a possible accretion disc. This was done with the internal total angular momentum of the sink particle (which is tracked in AMUSE) and the mass accreted by the sink particle in each time step. The size of the accretion disc with a certain mass $M$  and an angular momentum $L$ around a black hole with a mass of $M_{\bullet}$ when assuming a Keplerian orbit is then estimated with:

\begin{equation}
r_{\mathrm{disc}} = \frac{L^2}{GM^2M_{\bullet}}
\end{equation}

For $t=0$ or phases where no accretion is happening, we set $r_{\mathrm{disc}}$ manually to 0 as this would cause a division by $0$ otherwise. 

For different sizes of sink particles we computed the average size of the accretion disk. The result is shown in the middle panel of Figure \ref{fig:racc}. The size of the accretion disk rises when increasing the accretion radius and therefore increasing the mass of the accreted gas. The curve shows a maximum of a disc size of 100 - 200 au and decrease again at very large sink particle radii. Another estimation for the ideal accretion radius can be seen in right panel of Figure \ref{fig:racc} where we plot the radial distance of each star as a function of time. The closest approach of 100 au is reached by S14 in 2047. The accretion disc should not be larger than the closest approach of the S-stars. Considering all facts above we find that an accretion radius of $r_{\mathrm{acc}} = 100$ au seems to be the right choice. In this manner we have determined a self-consistent sink particle radius for our simulations, but in the results we present below the sensitivity with respect to this choice should be kept in mind.

\subsection{Escaping particles}

To reduce computational cost, we remove gas particles that are far away from the system and escaping in the absence of external forces.
We identify escaping particles using three criteria: 1) position - when the distance between the particle and the black hole is larger then twice the apocenter distance of the widest orbit. 2) velocity - when the velocity of the particle exceeds the local escape velocity. 3) direction - if the radial component of the velocity is positive, so the particle is moving away from the rest of the system.

%The accretion onto the black hole is complicated. As we are not interested in the properties of the accretion disk but only the total amount of gas that finally is accreted, we choose not to model the regions very close to the black hole. In the future it would be desired to couple our code with proper accretion disk models and add radiative transfer and feedback to the accretion process. In this work, we adopted a spherical sink particle with an accretion radius of $r_{\mathrm{acc}} = 0.05 r_{\mathrm{Bondi}}$ where $r_{\mathrm{Bondi}}$ is the Bondi-Hoyle accretion radius which is derived through:
%
%\begin{equation}
%r_{\mathrm{Bondi}} = \frac{2 G M_{\bullet}}{(v_w^2 + c_s^2)}
%\end{equation}
%
%where $v_w$ is the wind velocity and $c_s$ the sound speed which is dependent on the temperature of the gas. With wind velocities of around 1000 \kms, the sound speed is neglectable and the average accretion radius results in $r_{\mathrm{acc}} = 395$ au.

\begin{figure*}
  \centering  % this centres figure in column
     \includegraphics[width=\textwidth]{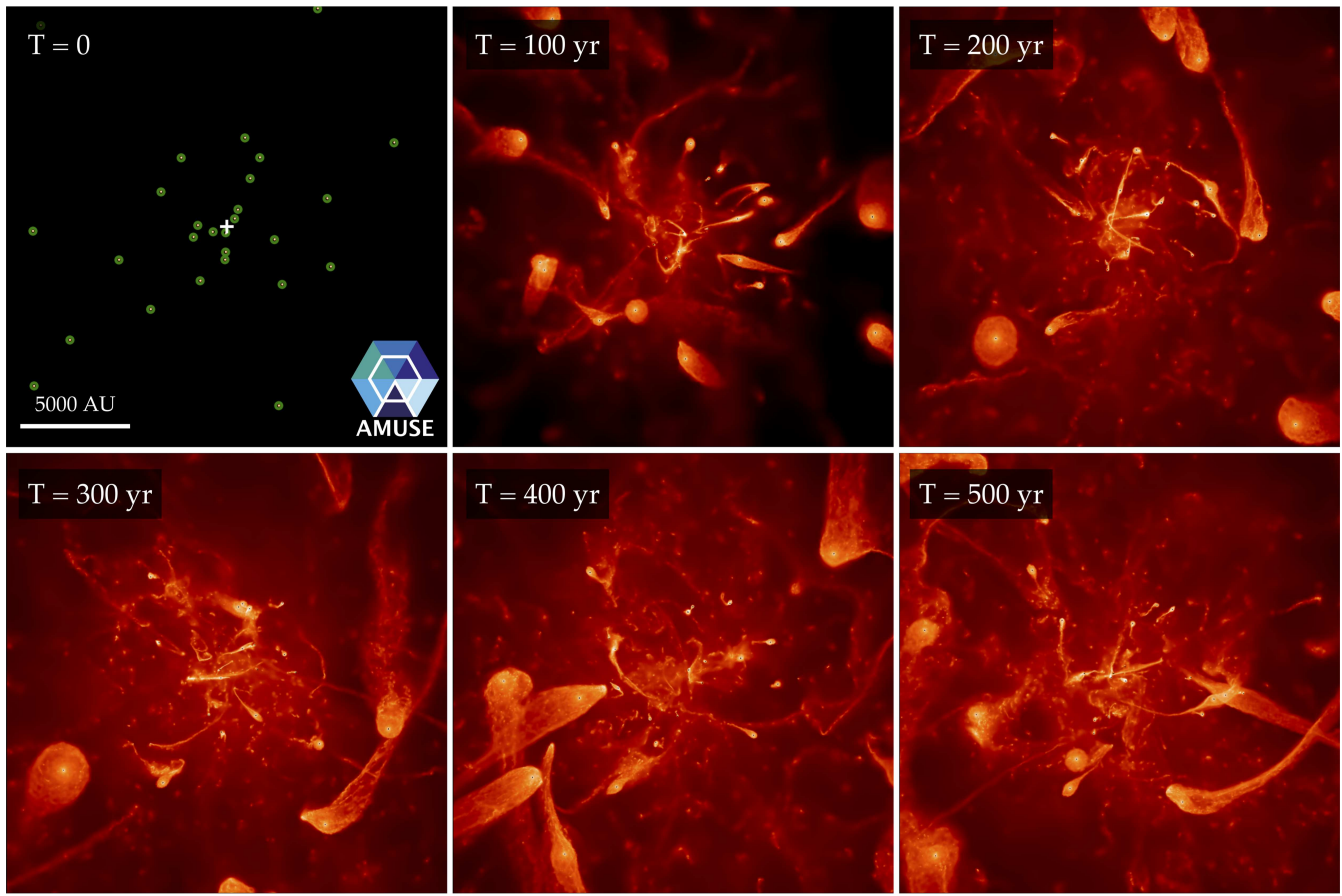}
  \caption{Snapshots at 6 different times in the simulation with all 27 stars. The images are centered on the black hole (white cross). The initial position of the stars are marked in the first panel with green circles. The animation of the simulation can be found at https://youtu.be/soGFgzRso3c.}
  \label{fig:snap}
\end{figure*}

\section{Results}\label{sec:res}

In this section we report on the results of the various tests that were performed with this code in order to verify its stability and convergence for different parameters. Understanding the behavior of the code for different number of particles and stars is crucial for understanding the results and their bias. In Figure \ref{fig:snap} we show a series of snapshots from the simulation with the 27 S-stars and the column density of the gas.  Note the noisy, grainy clumpy structures, visible especially in the later panels, result from the break up of the filamentary structure detaching from the winds due to ram pressure and gas piling up due to orbital changes.

\subsection{Cooling vs non-cooling}

The code written for this work connects stellar wind, hydrodynamics, stellar evolution and gravitational dynamics. It is used to simulate the S-star cluster in the galactic center. However, the code is open to the public and applicable to other astronomical problems. We therefore aimed at creating a general script that includes the major physical processes in this framework, including cooling. In this section we investigate the effect of the cooling for our specific scenario and argue that it is not needed for the galactic center.

The cooling was implemented in two distinct ways. The external way, i.e. an extra python routine that acts in the script and cools the gas at every time step using line cooling functions. And an internal method as it is implemented in \verb|Fi|. Both methods use slightly different cooling curves (the external cooling uses an approximation to the cooling curve used by the internal cooling). The different cooling curves and computational approaches might cause differences in the final output and therefore need to be investigated. We compare the results of simulations with and without cooling with respect to global properties and accretion rates in order to understand the importance of cooling. 

\begin{figure*}
  \centering  % this centres figure in column
  \includegraphics[width=0.45\textwidth]{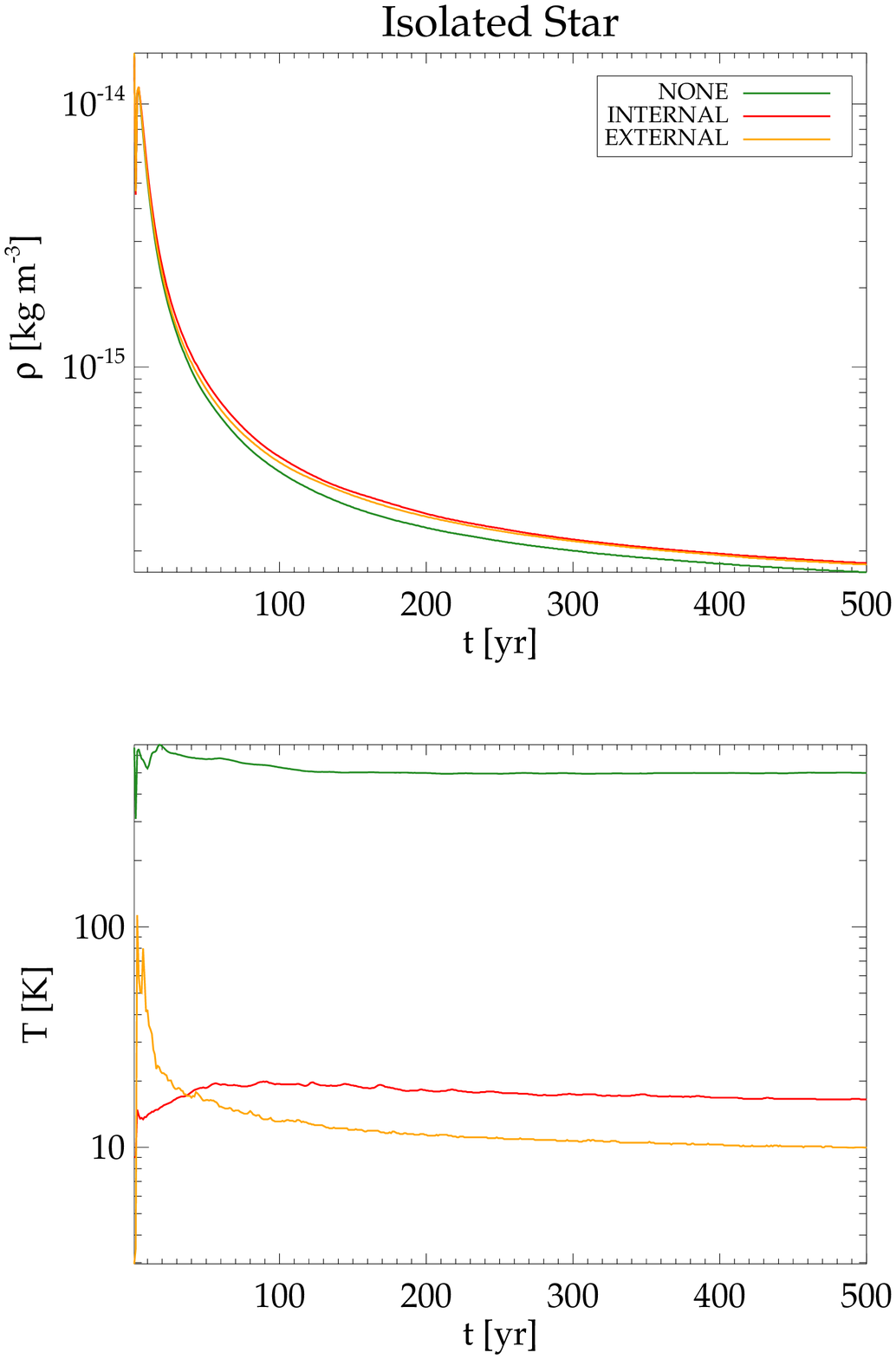}
  \includegraphics[width=0.45\textwidth]{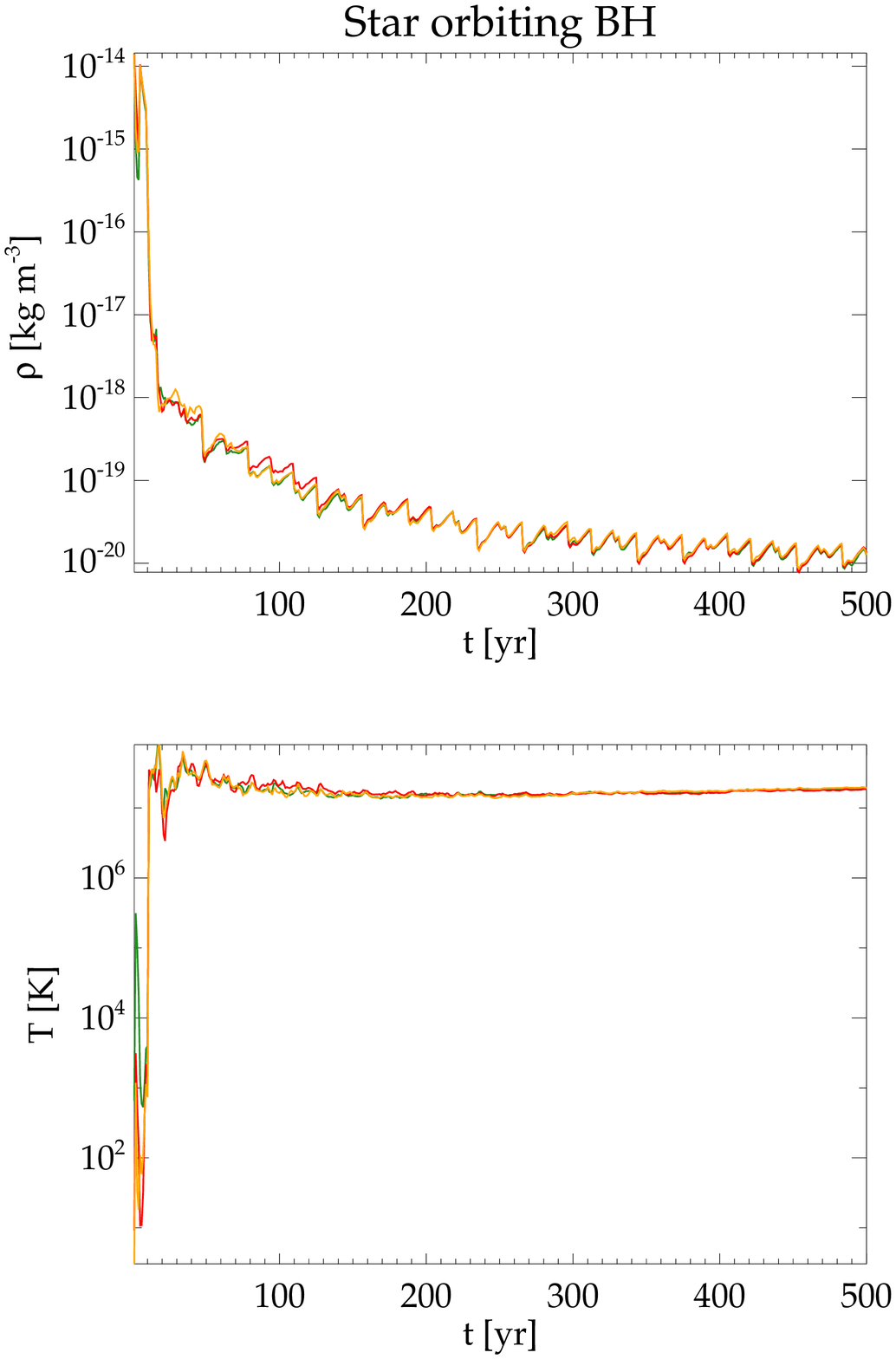}
  \caption{Temperature and density over time for the different cooling mechanism for an isolated star (left) and a star orbiting a black hole (right).}
  \label{fig:cool}
\end{figure*}

In the left panel of Figure \ref{fig:cool} we show the average temperature and density as a function of time for a simulation with no cooling and two simulations with internal and external cooling for a single isolated star. The right hand panel in Figure \ref{fig:cool} shows the same but with the star orbiting the black hole, allowing the stellar wind to collide with itself and to heat up through shocks. The figures show considerable differences in the temperatures and density only for the simulations with a single isolated star. For simulations with a star in motion around the black hole or multiple stars the cooling prescription adopted has no effect. The interactions between the stellar winds cause enough shock heating to overcome the cooling and maintain the temperature above $10^7 $K. At these temperatures the cooling times are on the order of $\sim 1 $ Myr, which is much longer than the dynamical timescales of the problem.

\subsection{Convergence}

Performing simulations with many particles is challenging and one has to find a good compromise between having a realistic simulations with as many (gas) particles as possible and performing the simulations in a reasonable time frame on computers that are available. Therefore it is crucial to perform convergence tests to find the minimum number of particles for which the physical properties of the system are not varying with particle number anymore. Our gas particle numbers are governed by the mass of individual particles ($M_{\mathrm{gas}}$). Since the mass loss of the stars remains the same throughout the simulation, a lower $M_{\mathrm{gas}}$ results in a higher number of particles that are emitted at every time step. 

\begin{figure}
  \centering  % this centres figure in column
     \includegraphics[width=0.49\textwidth]{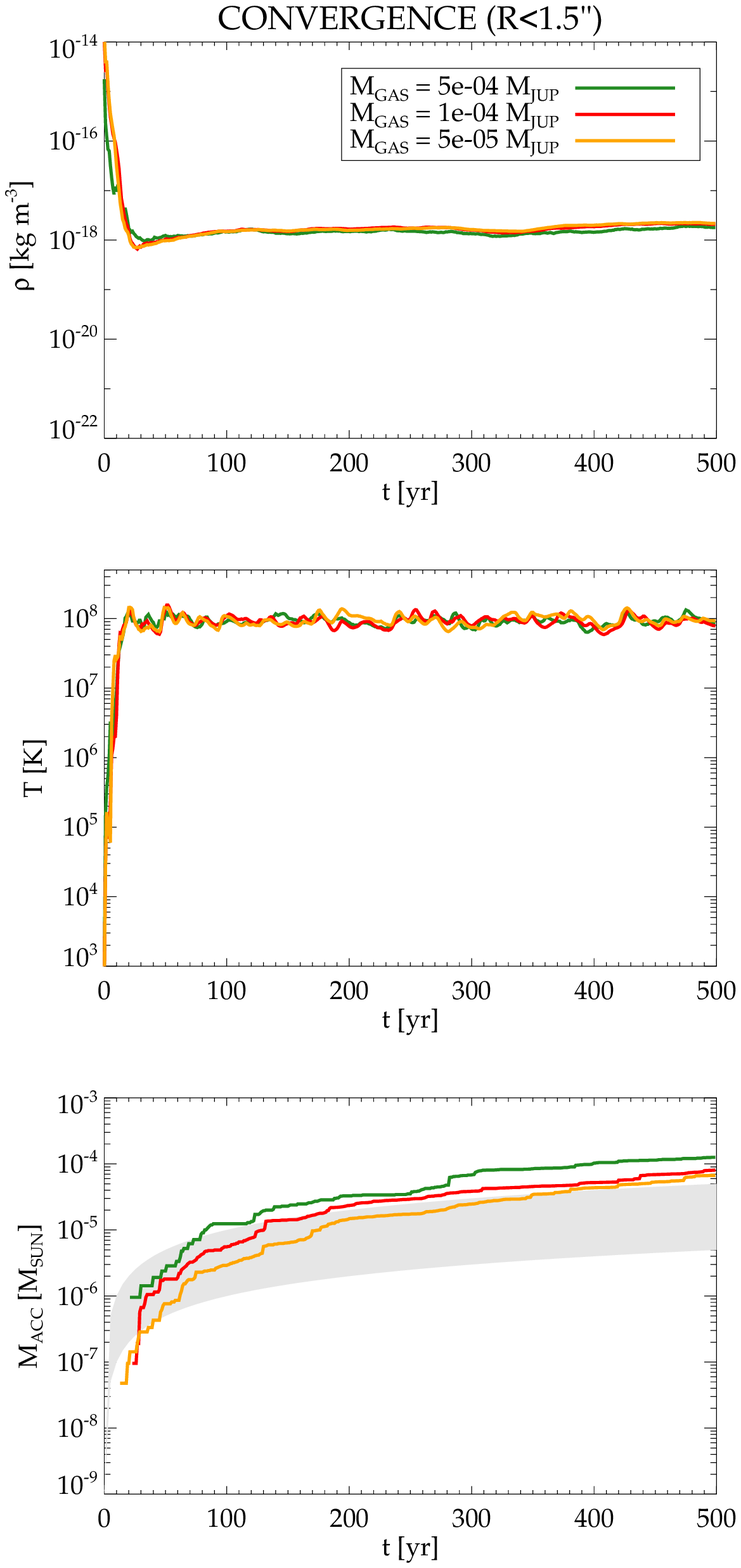}
  \caption{Overall properties of the stellar system with different gas particle numbers for an area around the black hole of the size of 1.5''. Shown are the evolution of the mean temperature (upper panel), the mean density (middle panel) and the accreted mass (lower panel). With the gray shaded area we mark the observed accretion rate from various sources.}
  \label{fig:all}
\end{figure}

Figure \ref{fig:all} shows the overall properties of the system (temperature, density and accretion) for three different values of particle masses: $M_{\mathrm{gas}} = 5 \times 10^{-4}, 10^{-4},$ and $5 \times 10^{-5} M_{\mathrm{Jup}}$ ($4.8 \times 10^{-7}, 9.5  \times 10^{-8}, 4.8 \times 10^{-8} \, M_{\odot}$) for a spherical area around the black hole with a radius of 1.5'' (12500 au). Note, that we are not stating particle numbers but rather particle masses due to the fact that our simulations produce new particles during every time step and therefore the particle number increases steadily, even though we remove escaped and accreted particles.
%\begin{figure}
%  \centering  % this centres figure in column
%  \includegraphics[width=0.49\textwidth]{stats_acc.eps}
%  \caption{Accretion rate for individual stars. The x-axis is in units of the end time of each run which differs for each star. The end time was set to a minimum of 500 yr and a maximum of three orbital periods. The orbital periods are marked with the grey lines. }
%  \label{fig:acc_all}
%\end{figure}

\begin{figure*}
  \centering  % this centres figure in column
  \includegraphics[width=\textwidth]{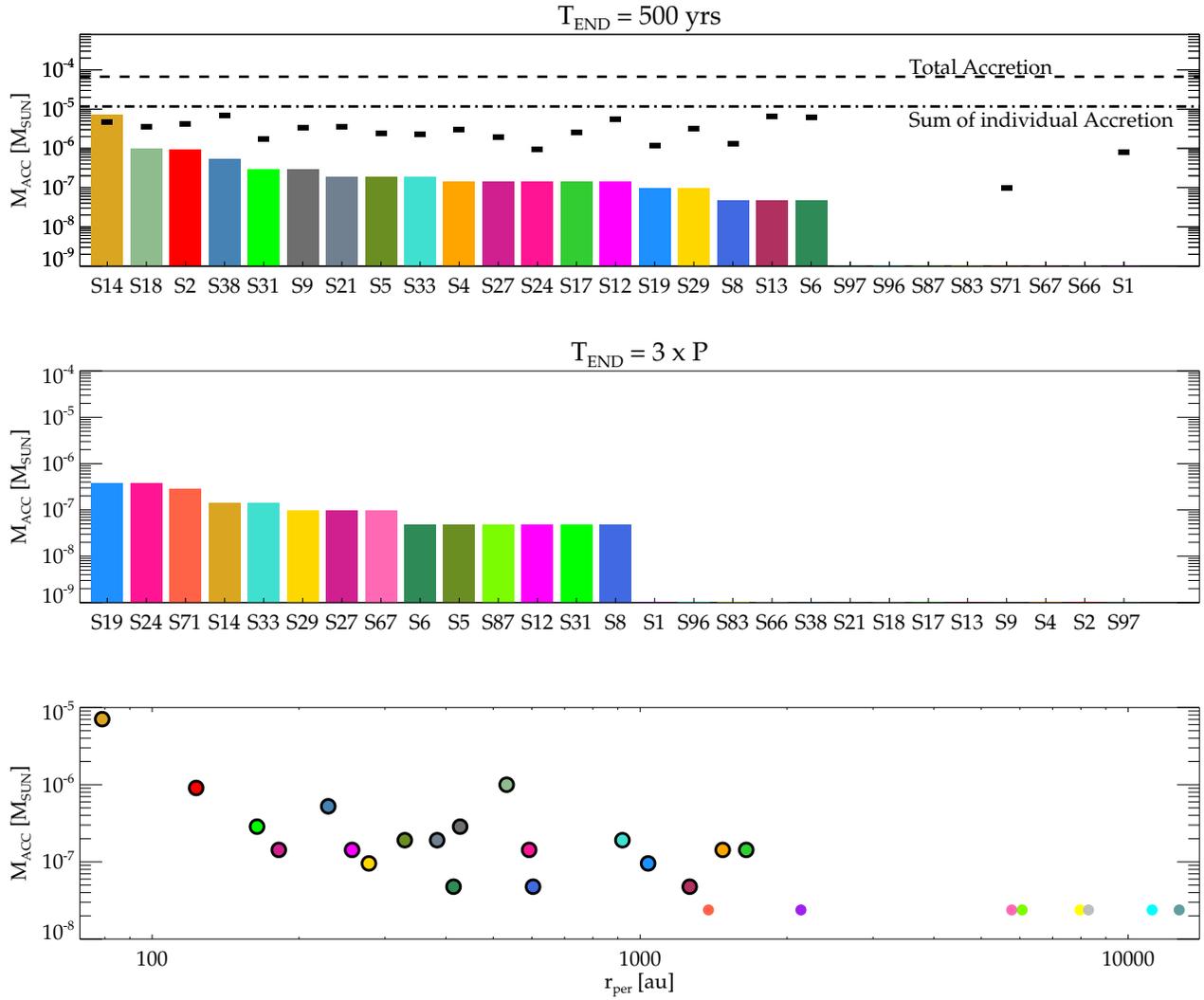}
  \caption{Upper panel: Total accreted gas mass for each individual star after after a total time of 500 yr. The total accretion (when all stars are part of the simulation) and the sum of individual accretion are represented by the dashed and dashed-dotted lines, respectively. Black bars mark the contribution of each individual star when all stars are part of the simulation. Middle panel: Total accreted gas mass for each individual star after completing three orbits. Lower panel: Total accretion after 500 years as a function of pericenter distance. Points without black borders mark stars that did not contribute any gas to the accretion yet (so we set the accretion artificially to a low value to keep them in the plot).} 
  \label{fig:acc_max}
\end{figure*}

Figure \ref{fig:all} shows no difference in temperature or density when considering different particle masses in a confined volume. The largest differences can be seen in the total accreted mass especially between the highest and lowest resolution model. However, these affects remain small and we consider our highest resolution model sufficiently well converged and we do not expect further major changes in the properties of the system when adding more particles.  

\subsection{Contribution of individual stars}

%Simulating individual stars separately has, apart from testing for convergence, another application. The accretion curves in Figure \ref{fig:acc_all} show the effect of the different orbits on the accretion flow onto the black hole. As a guideline, gray lines mark the moment of a pericenter passage. We recognize two different kinds of accretion flows. One that we call a step-like flow (e.g. S12, S14, S27, S29, S31,...) and one with a more continuous accretion flow but not always starting at the beginning of the simulations (e.g. S1, S4, S8, S13, S17,...). On the other hand, a case such as S66 also shows a step-like behavior, but can be explained by accretion of individual gas particles due to low resolution rather than a periodic accretion. Note that for the periodic step-like accretion flows the accretion burst agrees well with the pericenter passage of the star. Whether the accretion flow is step-like or continuous depends on the orbital parameters. It appears that a small pericenter distance is the key parameter: All stars with a pericenter $< 0.003$ pc show the step-like behavior while stars with larger pericenter distances exhibit the continuous accretion flow. In this case the gas forms an accretion disk and is continuously accreted by the black hole while short pericenter distances trigger shock-like accretions every time that star passes by. 

Simulating individual stars allows us to determine the accretion compared to a simulation containing all stars at once. The simulations were run for a minimum time of 500 years and a maximum of three times the star's orbital period in order to compare the accretion after completion of three orbits, and after for each star the same amount of time has passed. The longest simulation ran for 5100 years for the star S83 which has an orbital period of $\sim$ 1700 years. In Figure \ref{fig:acc_max} we show the accretion that is contributed by each star after 500 years (upper panel) and after passing three orbital periods (middle panel) as well as the accretion as a function of pericenter distance (lower panel). Also shown in dashed and dashed dotted lines is the total accretion when summing up the individual contributions and the total accretion that is obtained when running the simulation with all 27 stars at once, respectively.  Furthermore, the black bars mark the accretion of the individual star when run in the full simulation. The plot shows that the accretion is enhanced when letting the stars influence each other rather than run separately. This must be caused by the winds colliding and creating a violent and turbulent environments that leads to net angular momentum loss in the gas. This is also indicated by the contribution of the individual stars when simulated all together. The black bars mark the contribution of each star when they interact with each other. The order is very different from the order of most contributing stars in isolation. That strengthens the hypothesis that the accretion is enhanced through the interacting winds of the stars when simulated all together. Furthermore, the order of contributing stars changes, when normalizing by the orbital period (lower panel in Figure \ref{fig:acc_max}) which shows that the accretion rate for the stars in isolation is highly dependent on the orbital parameters of the stars. From the lower panel of Figure  \ref{fig:acc_max} one can see that the amount of accretion is roughly dependent on the pericenter distance but not perfectly correlated. Other factors such as eccentricity and period might also play a role.

From these plots we conclude that the order of the stars contributing most to the accretion is different when treating them separately or together in the simulations. For the separate case the stars S14, S18, S2, S38, S31, and S9 are the largest contributors (colored bars in the upper panel of Figure \ref{fig:acc_max}), while in the combined simulation the stars S83, S13, S6, S12, S14, and S2 contribute the most (black bars in the upper panel of Figure \ref{fig:acc_max}). From these figures we also see that a single star can not be responsible for the entire accretion and at least two S-stars are necessary to obtain the total accretion.

\subsection{Evolution of the orbital parameters}

\begin{figure*}
  \centering  % this centres figure in column
  \includegraphics[width=\textwidth]{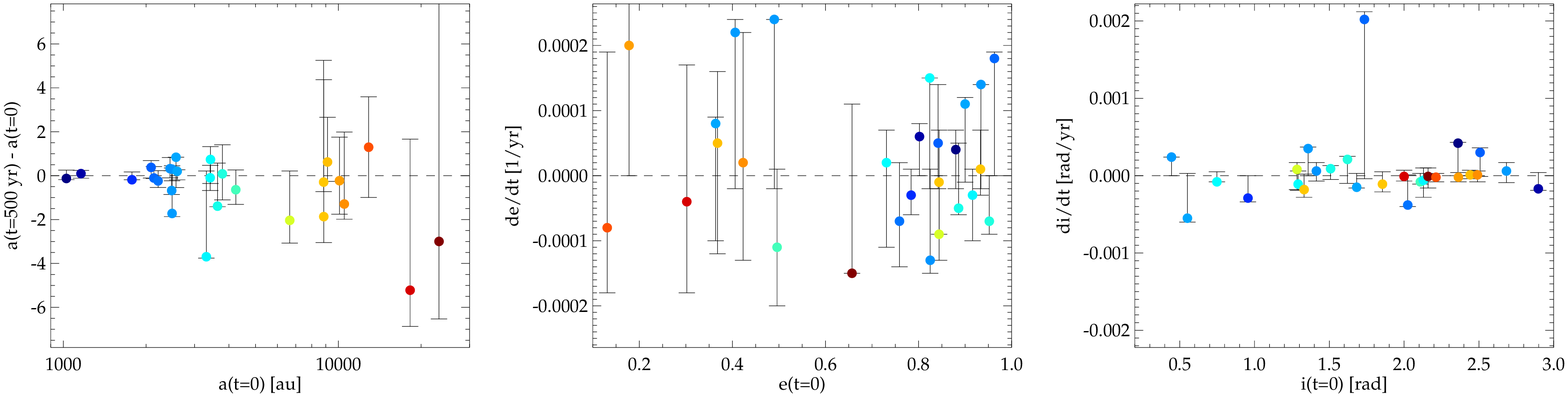}
  \caption{Change of orbital parameters during the simulation. Colors indicate the logarithmic length of the star's period, errorbars mark the maximum change in the parameter during its evolution.}
  \label{fig:orbital}
\end{figure*}

The mass loss of the stars, the wind momentum and the influence of the other stars are factors that can lead to variations in the stellar orbits. We calculate the orbital parameters from the star's position and velocity using the Kepler equations. In Figure \ref{fig:orbital} we show the final derivatives of the orbital parameters as a function of the start parameters for the semi-major axis (a), the eccentricity (e) and the inclination (i). To guide the eye, we color code the individual points with their logarithmic orbital periods. Dark blue indicate small orbital periods (i.e. the darkest point indicates S2) while dark red points have very long periods. The errorbars indicate the maximum changes/fluctuations in the parameters during the simulation. We do not see any correlation between length of period and change of orbital parameters. The overall changes in the parameters of all stars are rather small and probably solely due to star-star interaction as we do not see any difference in the orbital parameters when running the same simulation without the gas. This is expected as the total gas mass is negligible compared to the stellar mass.

%For stars with more than three apocenter passages we determine the change in orbital parameters by calculating the time derivative using a three-point (quadratic) Lagrangian interpolation. For stars with one or two apocenter passages we just calculate the change of the parameters from the beginning with respect to the time of the passage. Parameter changes for stars with no apocenter passage during the simulation were derived by the difference of the start and the end values. However, these results do not have much meaning since the orbital parameters can only be measure accurately at the star's apocenter.

\section{Comparison with observations}\label{sec:obs}

The estimates and upper limits for the accretion rate onto SgrA$^{\ast}$ are stated in Section \ref{sec:intro}. From these references we adopt a value of $10^{-7} - 10^{-8} M_{\odot}/\mathrm{yr}$. For densities and temperatures we use results from X-ray observations from the center of the Galaxy \citep[][]{baganoff_2003,bower_2003,nayakshin_2005}.

\subsection{Total accretion rate}

\begin{figure}
  \centering  % this centres figure in column
  \includegraphics[width=0.49\textwidth]{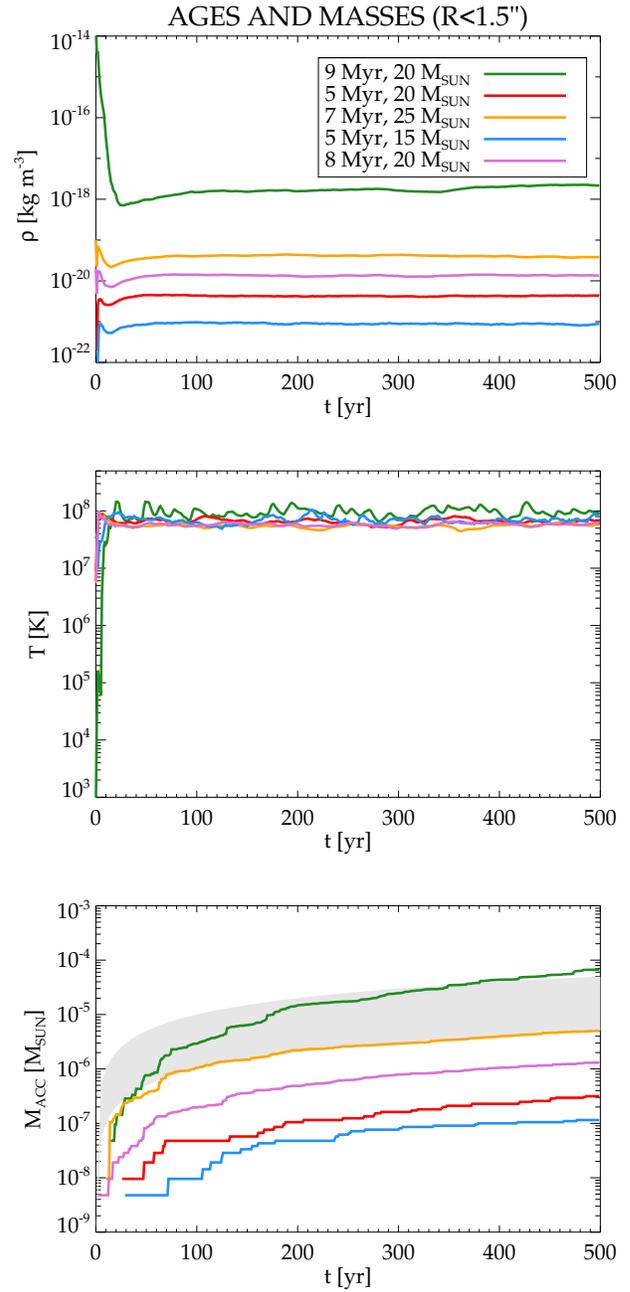}
  \caption{Same plot as in Figure \ref{fig:all} but for simulations with different ages and masses for the S-stars. Shown are the evolution of the mean temperature (upper panel), the mean density (middle panel) and the accreted mass (lower panel). With the gray shaded area we mark the accreted mass that would result from the observed accretion rate.}
  \label{fig:massage}
\end{figure}

The accretion rate we find in Figure  \ref{fig:all} is marginally consistent with the one observed, being close to the upper limits derived from observations \citep[grey shaded area][]{baganoff_2003,bower_2003,nayakshin_2005}. This can be caused by overestimating the mass loss rate of the stars or not treating feedback and accretion onto the black hole properly. For the latter we would need to extend our simulations to include radiative feedback, but this introduces a high computational cost. The mass loss rate of the stars, however, can be varied by varying the parameters of the stars in our code. We therefore rerun the simulations using different masses and ages of the stars. Figure \ref{fig:massage} shows the results of these additional runs. We found that the total accretion rate is sensitive to the mass loss rate, and thus in principle we can use the simulations to constrain the massloss rate - and therefore the ages and masses of the stars. Table \ref{tab:para} summarizes the variations in the stellar parameters that were applied.

\begin{figure}
  \centering  % this centres figure in column
  \includegraphics[width=0.49\textwidth]{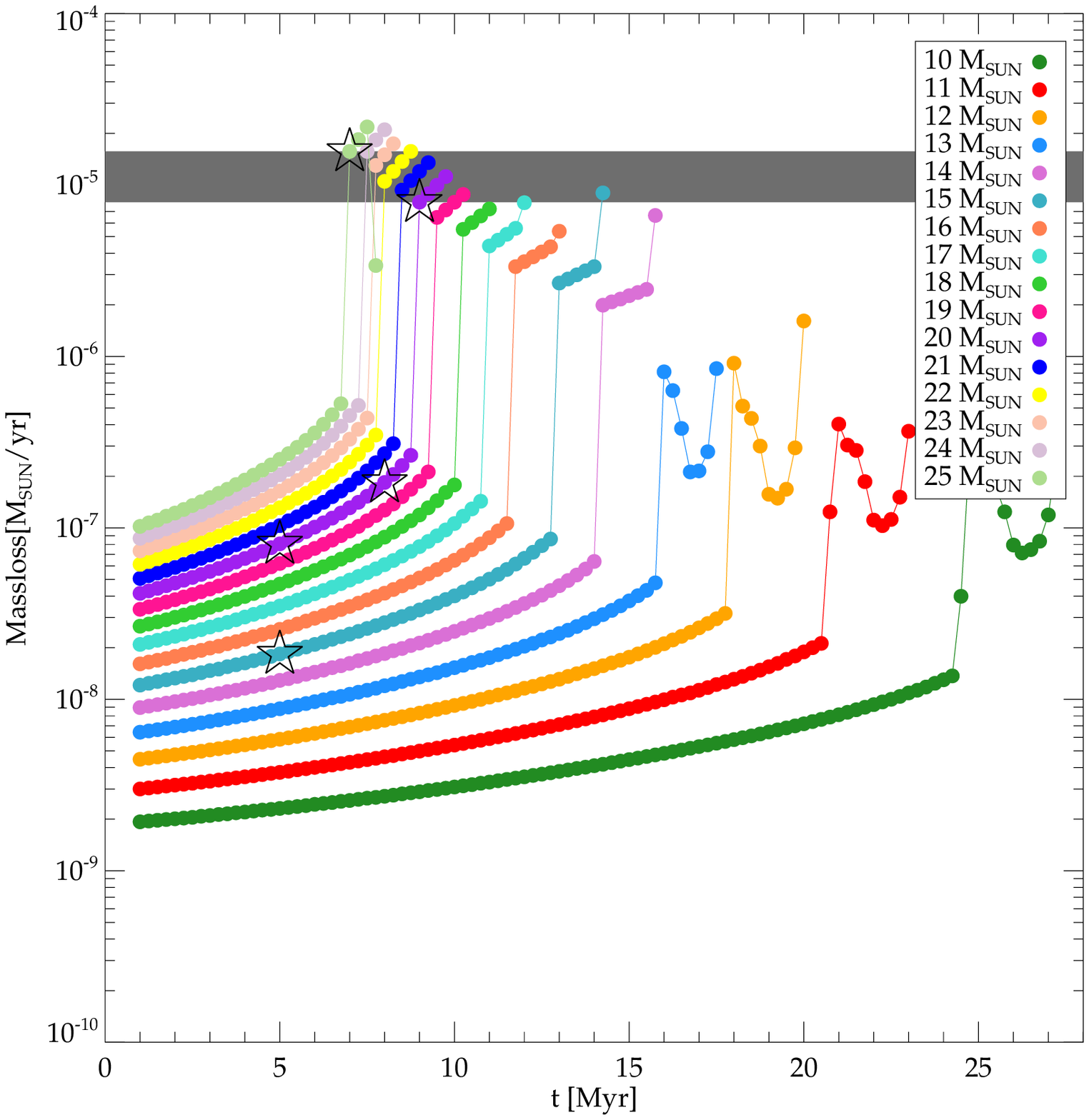}
  \caption{Mass loss rate as a function of age and mass of the star. The grey shaded area marks the mass loss rate that results in the correct accretion rate onto the black hole. This marks the range of ages and masses of the stars that reproduce the observations best. The star-symbols mark the position of the stars in our simulations (as listed in Table \ref{tab:para}.} 
  \label{fig:massloss}
\end{figure}

The accretion rate predicted by observations (grey shaded area in Figure \ref{fig:massage}) is traced by the simulation with $20\, M_{\odot}$ stars with an age of 9 Myr as an upper limit and the simulation with 7 Myr old stars that have masses $25\, M_{\odot}$ as a lower limit. The corresponding mass loss rate of this star is of the order of $10^{-5} M_{\odot}/\mathrm{yr}$. To connect masses and ages of the stars to their mass loss rate, we plot the mass loss rate of the stellar evolution code \verb|SSE| as a function of age for different masses in Figure \ref{fig:massloss}. To guide the eye we add shaded area of a mass loss rate between $8 \times 10^{-6} M_{\odot}/\mathrm{yr}$ and $10^{-5} M_{\odot}/\mathrm{yr}$ which we find, reproduces the observations well. We note that this restriction is only applicable to the S-stars contributing significantly to the total accretion rate such as S83, S13, S6, S12, S14, and S2 (see Figure \ref{fig:acc_max}).

However, these massloss rates are in contrast with \citet{martins_2008} who found a conservative upper limit of $3 \times 10^{-7} M_{\odot}/\mathrm{yr}$ for S02. Furthermore, as seen in Table \ref{tab:para} the two models that supply efficient amount to explain the accretion rate of the black hole, both require very high-mass stars with extreme parameters, (e.g. high luminosities) which are inconsistent with observations. S02 being the brightest of the S-stars with a K-magnitude of $\sim 14$ sets a limit to the maximal observed luminosity of $< 35 000 L_{\odot}$  \citep{martins_2008}. This excludes three of the five models in Table \ref{tab:para}.

%To connect masses and ages of the stars to their mass loss rate, we plot the mass loss rate of the stellar evolution code \verb|SSE| as a function of age for different masses in Figure \ref{fig:massloss}. To guide the eye we add shaded area of a mass loss rate between $8 \times 10^{-6} M_{\odot}/\mathrm{yr}$ and $10^{-5} M_{\odot}/\mathrm{yr}$ which we find, reproduces the observations well. The figure shows that our previous values for the stars ($M = 20 \, M_{\odot},$ age $= 9 \, \mathrm{Myr}$) lies on the upper end of the observed accretion range. The figure also allows us to make an estimate on the mass and age of the S-stars to explain the accretion rate that is observed. We find that the S-stars most probably have masses of $M = 19 - 25 \, M_{\odot}$ and ages of 7-10 Myr. Which is in good agreement with estimates from previous works. We note that this restriction is only applicable to the S-stars contributing significantly to the total accretion rate such as S83, S13, S6, S12, S14, and S2 (see Figure \ref{fig:acc_max}). As we did not run simulations with only subsystems of stars we do not know how this changes when only 2 or 3 stars show strong winds. We therefore assume that at least two of the S-stars need to be in this mass and age range to maintain the accretion rate.

\begin{table*}
\caption{Stellar parameters (of all 27 S-Stars), mass loss rates, and accretion rates for the different AMUSE simulations. }             % title of Table
\label{tab:para}      % is used to refer this table in the text
\centering
\begin{tabular}{cccccrcc}
\hline \hline
\noalign{\smallskip}
$M_{\ast}$       & Age    & $\log T_{eff} $    & $\log L$                   & Type  & $V_{\mathrm{wind}}$ & $\dot M_{\mathrm{wind}}$ &  $\dot M_{\mathrm{acc}}$ \\
 $[M_{\odot}]$ & [Myr]  & [$\log$ K]           &  $[\log L_{\odot}]$    &           &  [km/s]                        & $[M_{\odot}/yr]$                 &  $[M_{\odot}/yr]$               \\
 \noalign{\smallskip}
\hline
\noalign{\smallskip}
$  20.0$ & $   9.0$ &  $   3.56$ &  $   5.10$ &          Core Helium Burning & $     80.6$ & $    7.94e-06$& $    1.89e-07$\\
$  20.0$ & $   5.0$ &  $   4.49$ &  $   4.81$ &        Main Sequence & $   2380.7$ & $    8.07e-08$& $    6.68e-10$\\
$  25.0$ & $   7.0$ &  $   3.57$ &  $   5.28$ &          Core Helium Burning & $     83.0$ & $    1.57e-05$& $    2.49e-07$\\
$  15.0$ & $   5.0$ &  $   4.46$ &  $   4.39$ &        Main Sequence & $   2478.0$ & $    1.84e-08$& $    2.64e-10$\\
$  20.0$ & $   8.0$ &  $   4.42$ &  $   4.95$ &        Main Sequence & $   1799.6$ & $    1.84e-07$& $    3.28e-09$\\
\noalign{\smallskip}
\hline 
\end{tabular} 
\end{table*}

\subsection{Temperature and density}

Measuring temperatures in the galactic center requires high quality X-ray observations. \citet{wang_2013} performed 3 Ms of Chandra HETG gratings observations of Sgr A$^{\ast}$ and analyzed the 0th order spectrum to extract the properties of the central arcseconds of our Galaxy. Their best-fit temperature using a single-temperature model results in a temperature of $ k_B T \sim 3.5 \, (3.0, 4.0) \, \mathrm{keV}$ (90\% confidence interval) and an electron number density of $n \sim 110 \, \mathrm{cm}^{-3}$ (private communication with Frederick Baganoff). The density depends mainly on the temperature of the cooler surrounding gas, which is not tightly constrained due to the high absorption column to the Galactic Center.  The uncertainty is estimated by a factor of 2.  The quiescent spectrum was extracted from a circular radius of 1.5'' around the black hole.

From Figure \ref{fig:massage} (densities and temperatures of the gas within a circular radius of 1.5'' i.e. 12500 au) we find the average temperature to be $T \sim 5 \times 10^7 $ K and the mass density for stars that are between 7 and 9  Myr old and 20 - 25  $\msun$ is of the order of $10^{-20} - 10^{-18} \mathrm{kg} \, \mathrm{m}^{-3}$. The temperature corresponds to $k_B T \sim 4.3\, \mathrm{keV}$. This is somewhat higher than observed but still within the same order of magnitude. For the densities we divide our mass densities by the molecular weight $\mu = m_p/ (X \cdot (1.0+x_{\mathrm{ion}}) + Y \cdot (1.0+2.0 \cdot x_{\mathrm{ion}})/4.0 + Z \cdot x_{\mathrm{ion}}/2.0)$ with the proton mass $m_p$. $X=0.73$, $Y=0.25$, and $Z=0.02$  mass fractions of Hydrogen, of Helium, and of metals, respectively, and $x_{\mathrm{ion}} = 0.1$ is the ionization fraction ($0 < x_{\mathrm{ion}} < 1$), 1 means fully ionized. This transforms our mass density of $\rho \sim 10^{-20} - 10^{-18} \mathrm{kg} \,\mathrm{m}^{-3}$ to a number density of $n \sim 5 - 50 \, \mathrm{cm}^{-3}$. This value is smaller than the observed value especially when considering that the high density models are ruled out due to the stellar parameters in Section \ref{sec:res} and the densities for stellar parameters that are in agreement with the observations are as low as $n \sim 0.5 \, \mathrm{cm}^{-3}$ . However, due to its high model dependence, the measured value of the density is still very uncertain. Considering the uncertainty of a factor of 2, the discrepancy between the theoretical and measured value reduces to one order of magnitude. The measurements may be improved in the future by using more complex analysis and instruments with higher resolution, but the uncertainties depending on the temperature of the surrounding gas will remain a problem.

\section{Summary and conclusions}\label{sec:concl}

We use the Astrophysical Multi-purpose Software Environment (AMUSE) to simulate the accretion of stellar wind originating from the S-stars in the galactic center onto the supermassive black hole. Using the stellar evolution code \verb|SSE|, the gravitational dynamics code \verb|ph4|, hydrodynamics (SPH) code \verb|Fi|, and a \verb|bridge|-like coupling of gas and gravity we study how much of the gas is accreted onto a spherical sink particle with an accretion radius of $r_{\mathrm{acc}} = 100$ au. We find that a) the total accretion is lower when summing up the contributions of individual stars than when including all stars in the simulation, b) cooling is inefficient and the temperatures are high due to wind shocks, and c) the accretion rate is highly dependent on the stellar parameters. From point c) we are able to predict the stellar parameters of the S-stars to $M = 19 - 25 M_{\odot}$ and 7-10 Myr and explain the observed accretion rate without the need of the surrounding O-stars. The stars with the highest contribution to the accretion are S83, S13, S6, S12, S14, and S2. 

We note that the order of the most contributing star changes when simulating the stars in isolation. As a result we cannot determine which star is the largest contributor and how many stars are necessary to obtain these accretions rates. However, we know that one star is not enough and all stars simulated together fulfill this criterion. Therefore, it can be stated that at least two, but probably more, of the S-stars have to maintain copious winds and consequently have to be in their late evolutionary phase. This, however, is in contrast with observational facts that state that the S-stars are supposedly young B-type main sequence stars. The predicted luminosity and massloss rates do not agree with observations obtained by \citet{martins_2008}. We therefore conclude that the S-stars in their present stage are not the main contributors to the accretion rate of Sgr A$\ast$ and the inflow of gas from the massive O-stars is needed. However, it is most likely that the hot gas produced by the interacting winds of the S-stars might have an influence on the steady gas inflow from the O-stars and might even cause the outflow, observed by \citet{wang_2013}. This would be a very interesting topic for future work. The interaction of the two stellar wind accretion components is clearly highly complex and challenging to model, because of the large range in length and timescales involved.

When comparing temperatures and densities to recent X-ray observations, we find the temperatures satisfactorily reproduced by our simulations, but the densities are underestimated by at least one order of magnitude. The high uncertainty in the density measurements due to the dependence on the temperature of the surrounding gas, however, makes the comparison of measured and theoretical densities rather difficult. 

A main weakness of the simulations presented in this work lies in the description of the accretion onto the black hole. By using a spherical sink particle without radiative transfer, the accretion and the accretion disk close to the black hole are highly simplified. Furthermore, an 'empty' sink particle such as we are using creates an under-pressure in a region of higher density and artificially enhances the accretion. Future simulations using an accretion through sink particles should use a pressurized sink particle. Furthermore it would be interesting to see the effect of additional compact bodies, such as a stellar or an intermediate mass black hole \citep{merritt_2009}, on the accretion rate. 

%Another application for this code would be to predict accretion rates of intermediate-mass black holes in globular clusters. Ongoing studies of X-ray and radio emissions in the centers of globular clusters have resulted in stringent upper limits on the black-hole mass in case the black hole is accreting. 
%However, it is unclear how much gas is available for the black hole to accrete and how regular this accretion would be. Simulating a globular cluster using the AMUSE code presented in this work and assigning stellar winds to all stars in the cluster would give an estimate of the gas that can be contained in the cluster and that reaches the center for the black hole to accrete. 

\section*{Acknowledgments}

We thank Rainer Schoedel, Stefan Gillessen, Frederick Baganoff, Nathan de Vries and Arjen van Elteren for valuable information and discussion. Furthermore we thank the anonymous referee for valuable comments to improve the quality of this manuscript. This work was supported by the Nederlandse Onderzoekschool voor Astronomie (NOVA), the Netherlands Research Council NWO (grants \verb|#639.073.803| [VICI], \verb|#614.061.608| [AMUSE] and \verb|#612.071.305| [LGM]). Furthermore, we thank the anonymous referee for constructive comments that improved the manuscript.

\bibliographystyle{mn2e}
\bibliography{ref}

\label{lastpage}

\end{document}